\documentclass[a4paper,conference]{IEEEtran}
\IEEEoverridecommandlockouts
\usepackage{cite}
\usepackage{amsmath,amssymb,amsfonts}
\usepackage{algorithmic}
\usepackage{graphicx}
\usepackage{subfigure}
\usepackage{textcomp}
\usepackage{xcolor}
\usepackage{epstopdf}
\usepackage{underscore}
\usepackage{textcomp}
\usepackage{xcolor}
\usepackage{ulem}
\usepackage{epstopdf}
\usepackage{lettrine}
\usepackage{bm}
\usepackage[marginal]{footmisc}
\usepackage{multirow}
\usepackage{booktabs}
\usepackage{array}
\def\BibTeX{{\rm B\kern-.05em{\sc i\kern-.025em b}\kern-.08em
    T\kern-.1667em\lower.7ex\hbox{E}\kern-.125emX}}
\begin{document}

\title{A Signal Detection Scheme Based on Deep Learning in OFDM Systems\\
}

\author{\IEEEauthorblockN{Guangliang Pan\IEEEauthorrefmark{1}\IEEEauthorrefmark{2}, Zitong Liu\IEEEauthorrefmark{1}\IEEEauthorrefmark{2}, Wei Wang\IEEEauthorrefmark{1}\IEEEauthorrefmark{2}, Minglei Li\IEEEauthorrefmark{3}}\\
	 \IEEEauthorblockA{\IEEEauthorrefmark{1} College of Electronic and Information Engineering, Nanjing University of Aeronautics\\ and Astronautics, Nanjing, 211106, China.\\
	 \IEEEauthorrefmark{2}Key Laboratory of Dynamic Cognitive System of Electromagnetic Spectrum Space, Ministry of Industry \\and Information Technology, Nanjing, 211106, China. \\
	 \IEEEauthorrefmark{3}College of Control Science and Engineering, China University of Petroleum (East China), Qingdao, 266580, China. \\	
}
Email: glpan2020@nuaa.edu.cn, zt_liu@126.com, wei_wang@nuaa.edu.cn, liminglei977@gmail.com 
}
\maketitle

\begin{abstract}
Channel estimation and signal detection are essential steps to ensure the quality of end-to-end communication in orthogonal frequency-division multiplexing (OFDM) systems. In this paper, we develop a DDLSD approach, i.e., \uline{D}ata-driven \uline{D}eep \uline{L}earning for \uline{S}ignal \uline{D}etection in OFDM systems. First, the OFDM system model is established. Then, the long short-term memory (LSTM) is introduced into the OFDM system model. Wireless channel data is generated through simulation, the pre-processed time series feature information is input into the LSTM to complete the offline training. Finally, the trained model is used for online recovery of transmitted signal. The difference between this scheme and existing OFDM receiver is that explicit estimated channel state information (CSI) is transformed into invisible estimated CSI, and the transmit symbol is directly restored. Simulation results show that the DDLSD scheme outperforms the existing traditional methods in terms of improving channel estimation and signal detection performance. 
\end{abstract}


\section{Introduction}
Orthogonal frequency-division multiplexing (OFDM) is a multi-carrier wireless communication technology in cognitive radio. The OFDM adopts the mode of parallel transmission, which can effectively counter intersymbol interference (ISI) \cite{1998Robust,wu2014cognitive}. Meanwhile, the  presence of cyclic prefix (CP) has contributed to fight against ISI \cite{raghavendra2005exploiting}. Despite OFDM has many merits, it also suffers from multipath effects and other disturbances (e.g., doppler shift \cite{zhao1996sensitivity}, high peak average power ratio (PAPR) \cite{wulich2005definition}), which brings certain challenges to receiver signal recovery. Channel state information (CSI) plays a key role in solving these problems. Both coherence detection and demodulation require the support of CSI in OFDM systems. Generally, the CSI can be estimated by pilot before the signal detection of the receiver \cite{wu2013spatial}. With the estimated CSI, transmitted symbols can be recovered at the receiver \cite{ye2017power}. Only by adopting a suitable signal detection strategy for the OFDM wireless communication system, can the receiver detect OFDM signals with a lower bit error rate (BER) and thereby complete the whole high-quality signals recovery process \cite{sun2017millimeter,he2018deep,xu2018novel}. Nowadays, deep learning, a key technology for artificial intelligence (AI), has attracted the attention of many researchers, and it has achieved good practical application results in the fields of image processing and speech recognition  \cite{sejnowski2018deep,zou2019primer,zhang2020deep}. Meanwhile, it is also gradually expanding to the field of wireless communication, providing a preliminary reference solution to solve the problems in the field of wireless communication \cite{ding2018amateur}.

The research on signal detection has always been concerned by researchers. There are some conventional methods, such as least squares (LS) \cite{hamilton2011ofdm} and minimum mean-square error (MMSE) \cite{abdelgader2017channel}, have been widely used in channel estimation under different communication systems. Both LS and MMSE are non-blind channel estimation methods, which need the support of pilot sequence. Besides, at the receiving end, due to the presence of pilot sequence, the ideal channel estimation cannot be completed. Even if perfect channel estimation can be performed, channel compensation during signal demodulation amplifies the noise signal. In \cite{yecsilyurt2017hybrid}, for space-division multiple access OFDM (SDMA-OFDM) systems, the authors proposed hybrid maximum likelihood MMSE (ML-MMSE) adaptive multiuser detection based on joint channel estimation to ensure tradeoff between the complexity and BER performance. In \cite{pham2016channel}, the authors proposed a multistep channel estimation scheme that utilized pilot subcarriers and data estimates. Then, for signal detection, a high-performance bidirectional M-algorithm (BDMA) was proposed for trellis-based equalization. 

Deep learning is introduced into the physical layer, which provides an effective method to solve many problems in wireless communication, such as channel encoding and decoding \cite{liang2018iterative}, modulation recognition \cite{wang2019data}, channel estimation and detection \cite{amirabadi2020deep}. For channel estimation and signal detection, In \cite{ye2017power}, a deeper versions of artificial neural networks (ANNs) was used to counter nonlinear distortion and interference in wireless channel characteristics and frequency selectivity to realize end-to-end channel estimation and symbol detection. \cite{yi2020deep} proposed a channel estimation network (CENet) and a channel conditioned recovery network (CCRNet). The CENet was used to replace the traditional interpolation procedure, and tne CCRNet was used to recover the transmit signal. In \cite{xiang2020deep}, the authors used a two-layer neural network (TNN) and a deep neural network (DNN) to jointly design the pilot and channel estimator, and then used another DNN to complete iterative optimization in massive multiple inputs multiple outputs (MIMO) systems. In \cite{yang2019deep}, the authors used the DNN model to train the simulated data offline, and then made online estimation for the dual-channel. Meanwhile, a pre-training method was designed for DNN to further improve the performance of the algorithm. The above methods based on deep learning basically adopt the deep version of ANNs to complete channel estimation and signal detection, while for deep learning, there are other neural network models that are more conducive to mining sequence data features and improving the detection performance of the system.
 
To address these issues, we develop a DDLSD approach, i.e., \uline{D}ata-driven \uline{D}eep \uline{L}earning for \uline{S}ignal \uline{D}etection in OFDM systems. The main contributions of this paper are summarized as follows:

\begin{itemize}
	\item We propose a signal detection method based on data-driven, the long short-term memory (LSTM) neural network as a black box replaces the channel estimation, equalization and symbol detection process of OFDM systems, turning it into a single operation.
	\item We evaluate DDLSD method under various parameter configurations and our experiment results show that the BER of DDLSD method is significantly lower than that of LS and MMSE algorithms. Meanwhile, we find that the DDLSD is less sensitive to parameter changes, which indicates that the proposed DDLSD method has strong robustness.
\end{itemize}

The remainder of the paper is organized as follows. In Section~\ref{SEC2}, we introduce the OFDM system model. A data-driven signal detection method based on deep learning is proposed in Section ~\ref{SEC3}. The simulation results and analysis are shown in Section ~\ref{SEC4}. Finally, conclusion is provided in Section~\ref{SEC5}.

\section{The OFDM System Model}\label{SEC2}
\begin{figure}[htbp]
	\centerline{\includegraphics[height=46mm,width=70mm]{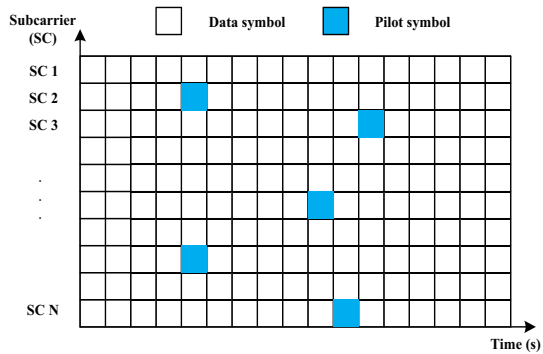}}
	\caption{The OFDM transmit symbol.}
	\label{inter}
\end{figure}
Based on Fig.~\ref{inter}, we consider an OFDM system with parallel transmission via $N$ subcarriers, where the transmitted symbols $\bm{{\rm D}}$ of sequence length $N$ consists of data symbols and pilot symbols $\bm{{\rm D}}= [D_0, D_1,..., D_{N-1}]^T$. The transmitted symbols are mapped to constellations by modulation (e.g., quadrature phase shift keying (QPSK) or 16 quadrature amplitude modulation (16QAM)). The modulation process can be implemented by using an $N$-point inverse FFT (IFFT) algorithm in OFDM systems, where its output during the $n$-th OFDM block can be written as $\bm{{\rm d}}(n) = \bm{{\rm A}}^I \bm{{\rm D}}(n) $, where $\bm{{\rm A}}$ is the normalized FFT matrix $\mathbb{C}^{N \times N}$, and hence, $\bm{{\rm A}}^I$ is the IFFT matrix \cite{saci2020direct}. Then, one bright spot is the insertion of a CP of length $N_{cp}$, no less than the channel maximum experimental spread ($C_h$), into the transmission signal. Therefore, the total length of OFDM symbols is $N_T=N+N_{cp}$ in continuous transmission for $T$ seconds. 

The OFDM transmit signal enters the front end of the receiver through wireless channel, the sampling period is $T_s=T/N_T$. The channel is considered to consist of $C_h + 1$ independent multipath components each of which has a gain $h_m \sim \mathcal{CN}(0,2 \sigma^{2}_{h_m})$ and delay $m \times T_s$, where $m \in \{0, 1,..., C_h\}$. So, the received signal is represented in the time domain as
\begin{equation}
\bm{{\rm y}} = \bm{{\rm h}} \otimes \bm{{\rm d}}+\bm{{\rm e}},
\end{equation}
where $\bm{{\rm h}}$ is channel matrix, $\otimes$ is cyclic convolution, $\bm{{\rm y}}$, $\bm{{\rm e}} \in \mathbb{C}^{N \times 1}$, $e \sim \mathcal{CN}(0,2 \sigma^{2}_{m})$ is the additive white Gaussian noise (AWGN). Then, the received sequence after removing the CP and applying the FFT is represented in the frequency domain as
\begin{equation}
\bm{{\rm Y}} = \bm{{\rm H}} \odot \bm{{\rm D}}+\bm{{\rm E}},
\end{equation}
where $\odot$ represents multiply element by element, $\bm{{\rm Y}}, \bm{{\rm H}}, \bm{{\rm D}}, \bm{{\rm E}}$ are the FFT of $\bm{{\rm y}}, \bm{{\rm h}}, \bm{{\rm d}}, \bm{{\rm e}}$, respectively.

The transmitted signal reaches the equalizer after FFT, and then carries out coherent maximum likelihood detector (MLD), which can be expressed as 
\begin{equation}\label{equ7}
\hat{\bm{{\rm D}}} = {\rm arg} \mathop {\rm min} \limits_{\tilde{\bm{{\rm D}}}} ||\bm{{\rm Y}}- \bm{{\rm H}}\tilde{\bm{{\rm D}}}||^2,
\end{equation}
where $\tilde{\bm{{\rm D}}} = [\tilde{ D_0}, \tilde{D_1}, ..., \tilde{ D_{N-1}}]$ represents the trial values of $ \bm{{\rm D}} $, and $||.||$ represents the Euclidean norm. According to equation~(\ref{equ7}), our purpose is to solve the optimal $\hat{\bm{{\rm D}}}$ and minimize its error with the actual signal $\bm{{\rm D}}$.

\section{Signal Detection Based on Deep Learning}\label{SEC3}
\begin{figure}[htbp]
	\centerline{
		\includegraphics[width=88mm,height=58mm]{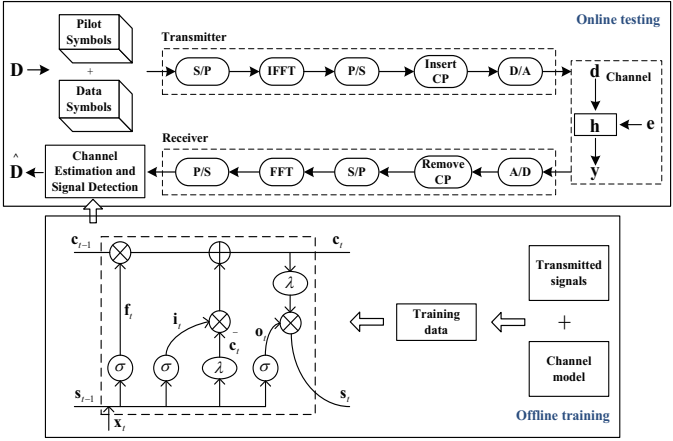}}
	\caption{The design of DDLSD method.}
	\label{fig8}
\end{figure}
In order to obtain the optimal $\hat{\bm{{\rm D}}}$, we propose a signal detection scheme based on deep learning. Fig.~\ref{fig8} illustrates how the DDLSD is implemented. Here, the LSTM network is designed to recover transmit signal. On the one hand, OFDM signals are time series with hidden features. On the other hand, the LSTM's unique memory ability in learning time series features is significantly better than that of ANN and convolutional neural network (CNN). Therefore, the LSTM network is considered to be applied in signal detection of OFDM systems. As can be seen from Fig.~\ref{fig8}, the implementation of DDLSD is divided into two parts: offline training and online testing. 

1) \textit {Offline training}

Simulation data obtained through OFDM system and wireless channel modeling are used as training data $\{(x_{1},y_{1}), ... ,(x_{n},y_{n})\}$, where $x_{n}$ represents the OFDM symbol, $y_{n}$ represents the label value. Specifically, the LSTM network has $L$ layers, and the first layer is denoted as the input layer. First, the OFDM training data through the forgetting gate of LSTM, which determines which information is retained in the output $\bm{{\rm s}}_{t-1}$ and unit state $\bm{{\rm c}}_{t-1}$ at the previous moment to the current moment. The input of the forgetting gate is the output $\bm{{\rm s}}_{t-1}$ of the previous moment and the input $\bm{{\rm x}}_t$ of the current moment, and the output through the forgetting gate in the LSTM can be expressed as follows \cite{huang2015bidirectional}
\begin{equation}\label{eq12}
\bm{{\rm f}}_t=\sigma(\bm{{\rm W}}_f[\bm{{\rm s}}_{t-1},\bm{{\rm x}}_t]+\bm{{\rm b}}_f),
\end{equation}
where $\sigma$ represents sigmoid activation function, $\bm{{\rm W}}_f$ and $\bm{{\rm b}}_f$ represent the weight and bias of the forgetting gate respectively. Then, OFDM data enters the input gate, which determines which information of the input $\bm{{\rm x}}_t$ at the current moment is retained to the current state unit $\bm{{\rm c}}_t$, and use the activation function to realize the update of the state unit $\bm{{\rm c}}_t$. The specific formula is expressed as follows
\begin{equation}\label{eq13}
\bm{{\rm i}}_t=\sigma(\bm{{\rm W}}_i[\bm{{\rm s}}_{t-1},\bm{{\rm x}}_t]+\bm{{\rm b}}_i),
\end{equation}
\begin{equation}\label{eq14}
\tilde{\bm{{\rm c}}}_t=\lambda(\bm{ {\rm W}}_c[\bm{{\rm s}}_{t-1},\bm{{\rm x}}_t]+\bm{{\rm b}}_c),
\end{equation}
where $\bm{{\rm W}}_i$ and $\bm{{\rm b}}_i$ represent the weight and bias of the input gate respectively. $\lambda$ represents tanh activation function. $\bm{{\rm W}}_c$ and $\bm{{\rm b}}_c$ represent the weights and biases of alternative update units, respectively. The updated cell state $\bm{{\rm c}}_t$ is
\begin{equation}\label{eq15}
\bm{{\rm c}}_t=\bm{{\rm f}}_t*\bm{{\rm c}}_{t-1}+\bm{{\rm i}}_t*\tilde{\bm{{\rm c}}}_t,
\end{equation}
where $*$ represents the dot product between elements. After information is selectively remembered and updated, it finally enters into the output gate. The formula is expressed as
\begin{equation}\label{eq16}
\bm{{\rm o}}_t=\sigma(\bm{{\rm W}}_o[\bm{{\rm s}}_{t-1},\bm{{\rm x}}_t]+\bm{{\rm b}}_o), \bm{{ \rm s}}_t = \bm{{\rm o}}_t * \lambda(\bm{{\rm c}}_t),
\end{equation}
where $\bm{{\rm W}}_o$ and $\bm{{\rm b}}_o$ represent the weight and bias of the output gate respectively. We adopt the cross-entropy algorithm to improve training speed, which is
\begin{equation}\label{eq18}
\psi = \sum_{i=1}^{p}[D_ilog(\hat{D_i})+(1-D_i)log(1-\hat{D_i})].
\end{equation}
where $p$ is the number of input units, $\hat{D_i}$ represents the estimated OFDM symbols, $ D_i $ represents the real OFDM symbols. Assuming that the layer $l$ has $M_l$ nodes, the overall cost of this model in the training process can be calculated as \cite{zhang2018spectrum}
\begin{equation}\label{eq19}
\Gamma(W,b) = \psi +\dfrac{\eta}{2}\sum_{l=1}^{L-1}\sum_{i=1}^{M_l}\sum_{j=1}^{M_{l+1}}(W_{ji}^{l})^2,
\end{equation}
where $W_{ji}^{l}$ represents the weight between the $i$ node in layer $l$ and the $j$ node in layer $l+1$ of the neural network and $\eta$ represents the attenuation coefficient. Let $\rm{U} = \{W,b\}$, the objective function is
\begin{equation}
\rm{U} = {\rm arg} \mathop {\rm min} \limits_{\rm{U}} \Gamma(W,b).
\end{equation}
In order to get the optimal $\rm{U}$, we use the gradient descent algorithm:
\begin{equation}\begin{split}\label{eq25}
W(s+1) \leftarrow W(s)-\alpha \dfrac{\partial \Gamma(W,b)}{\partial(W)},\\
b(s+1) \leftarrow b(s)-\alpha \dfrac{\partial \Gamma(W,b)}{\partial(b)},
\end{split}\end{equation}
where $W(s)$ and $b(s)$ denote weight and bias of $s$-th training, respectively. $\alpha$ is the learning rate.

2) \textit {Online testing}

After the training of the model, the trained model was used for online detection of OFDM signals. We input $N$ groups of test data into the trained model and get the output. To demonstrate the effectiveness of the proposed DDLSD method, Monte Carlo simulation is used to count the number of output values equal to the actual values, and then the BER is calculated as
\begin{equation}
P_{BER}= 1 -\dfrac{d}{N},
\end{equation}
where $d$ represents the statistic where the predicted value of the model is equal to the true value.

\section{Simulation Results and Analysis}\label{SEC4}
We have conducted several experiments to demonstrate the effectiveness of DDLSD for signal detection in OFDM wireless communication systems. Simulation parameters are shown in Table~\ref{table1}. The DDLSD and the traditional methods are tested online under different signal-to-noise ratios (SNRs) to compare their performance, and the BER is used for performance index. It can be seen from these experiments that the BER of the DDLSD in signal detection is significantly lower than other methods, which indicates that the DDLSD has stronger robustness. For an OFDM system, 64 subcarriers and CP with a length of 16 are used, and 2 OFDM blocks (1 OFDM block is composed of a set of data symbols and a set of pilot symbols) are used as the data set. The ratio of the train set and the validation set of the data set is 4:1. The conventional 3GPP TR38.901 channel is used as wireless channel model in OFDM system (For wireless channel model, other channel models can also be used, e.g., Riley decay channel).
\begin{table}[h]
	\centering
	\small
	\caption{Model Parameters.}\label{table1}
	\begin{tabular}{c c }
		\toprule[2pt]  
		Parameter& Value \\
		\midrule 
		Optimizer& Adam\\
		Gate Activation Function ($\sigma $)& Sigmoid\\
		State Activation Function ($\lambda $)& Tanh\\ 
		Input Size& 256\\
		MiniBatch Size& 1000\\
		MaxEpochs Size& 100\\
		Num Hidden Units& 16\\
		Initial Learn Rate ($\alpha $)& 0.01\\
		Learn Rate Drop Factor ($\eta$)& 0.1\\
		Gradient Threshold& 1\\
		\bottomrule[2pt]  
	\end{tabular}
\end{table}

\subsection{Impact of Pilot Numbers}
\begin{figure}[htbp]
	\centerline{
		\includegraphics[width=78mm,height=72mm]{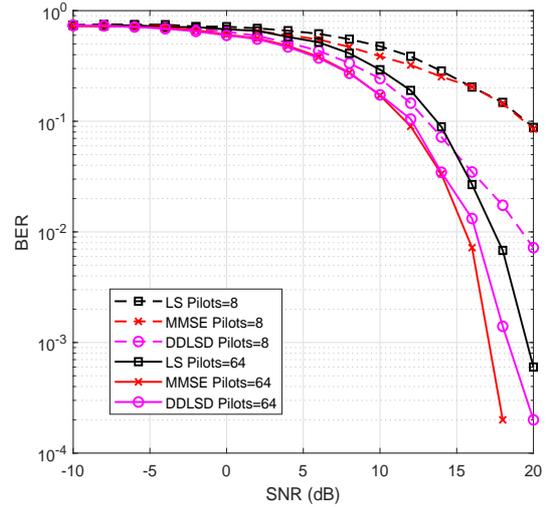}}
	\caption{BER performance under different pilot numbers.}
	\label{pilot}
\end{figure}

In this experiment, we analyze the influence of pilot numbers on the signal detection performance of DDLSD. The number of pilots is set as 8, 64, respectively. The performance curve of DDLSD is shown in Fig.~\ref{pilot}. When the number of pilots is 8, it can be seen from Fig.~\ref{pilot} that the BER curves of LS and MMSE algorithms are almost the same under different SNR. When the SNR changed from 10dB to 20dB, the detection performance of DDLSD improved significantly, while LS and MMSE did not change significantly. This indicates that DDLSD can quickly capture the characteristic information of OFDM signals.

From the Fig.~\ref{pilot}, when the number of pilots is 64, the performance of the traditional algorithm is equal to or even better than proposed DDLSD method. These changes indicate: (i) The performance of the LS and MMSE is greatly influenced by pilot, which is positive correlation; (ii) The LSTM is less sensitive to the change of pilot numbers. However, the detection performance of DDLSD does not fluctuate greatly due to the change of pilot numbers. This shows that DDLSD has certain robustness to the variation of the pilots. Similarly, as can be seen from Table~\ref{table:2}, when BER is $10^{-3}$, the detection accuracy of MMSE is better than DDLSD under the number of pilots is 64. This is because the increase of the pilot numbers reduces the characteristic information of OFDM signals, which reduces the judgment ability of the network. Besides, when the number of pilots is 64 and the SNR is 20dB, the BER of DDLSD is significantly lower than that of the method proposed in literature \cite{ye2017power}.

\begin{table}[h]
	\centering
	\small
	\caption{Comparison of BERs for different pilot numbers.}\label{table:2}
	\begin{tabular}{c c c c}  
		\toprule[2pt]
		Algorithm & Num-pilot &SNR \tiny {($BER=10^{-2}$)}& SNR \tiny {($BER=10^{-3}$)}\\ \hline
		\multirow{2}{*}{LS} & 8 & 20dB+ & 20dB+ \\
		& 64 & 17.5dB & 19.5dB \\ \hline
		\multirow{2}{*}{MMSE} & 8 & 20dB+ & 20dB+ \\
		& 64 & 15.6dB & 17dB \\ \hline
		\multirow{2}{*}{DDLSD} & 8 & 19.2dB & 20dB+ \\
		& 64 & 16.1dB & 17.9dB \\ 
		\bottomrule[2pt]
	\end{tabular}
\end{table}
\subsection{Impact of CP}
\begin{figure}[htbp]
	\centerline{
		\includegraphics[width=78mm,height=72mm]{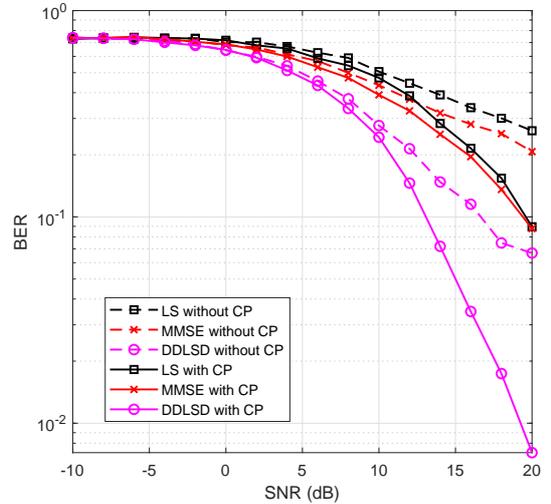}}
	\caption{Comparison of BER performance with and without CP.}
	\label{cp}
\end{figure}

The number of pilots is 8, QPSK modulation is adopted, and the SNR of training process is 20dB. We consider two situations: One is that no CP is inserted into OFDM signal, the another case is to insert CP into the OFDM signal. In these two cases, the performance comparisons between the DDLSD and the other two methods are presented in Fig.~\ref{cp}.

From the Fig.~\ref{cp}, neither LS nor MMSE can effectively estimate the channel. However, the BER of DDLSD under different SNR is lower than that of LS and MMSE. Besides, by observing the degree of BER variation of the methods, it can be found that the performance fluctuations of LS and MMSE are less affected by CP, while the performance fluctuations of DDLSD are more affected by CP when SNR is 20 dB. This result indicates that the DDLSD has learned the wireless channel characteristic information of OFDM system. Here, the computational complexity of traditional methods seems simple, especially LS. The complexity of DDLSD is mainly reflected in the training stage, and the complexity of the training stage is reflected in the time dimension. The trained DDLSD is directly used for signal detection, and the BER is significantly lower than the traditional method. Besides, for the time loss, we pay more attention to the accuracy of signal recovery. 

\subsection{Impact of Different Modulation Modes}
\begin{figure}[htbp]
	\centerline{
		\includegraphics[width=78mm,height=72mm]{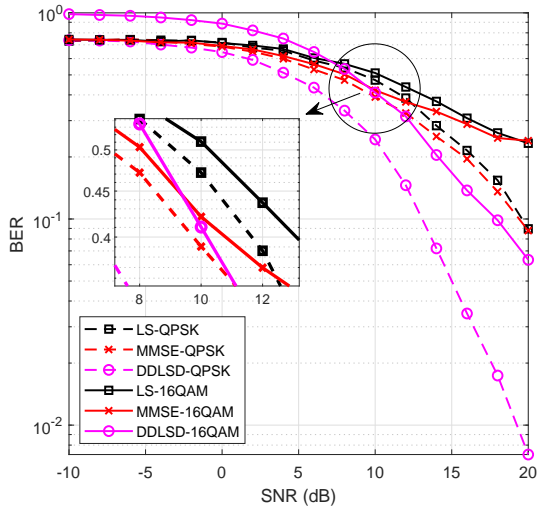}}
	\caption{BER curves of DDLSD and traditional algorithms in different modulation modes.}
	\label{mod}
\end{figure}

To illustrate the reliability and intelligence of the proposed DDLSD method, different modulation modes are considered. The same parameters are used, and the performance comparison of various methods is presented in Fig.~\ref{mod}. From the Fig.~\ref{mod}, the BER of DDLSD is significantly lower than LS and MMSE under QPSK and 16QAM. Table~\ref{table:3} shows that the SNR of DDLSD is lower than LS and MMSE algorithms under at the same BER. Especially in QPSK, the DDLSD has strong robustness to noise and maintains good signal detection performance in a harsh communication environment. The above show that the LSTM can still learn feature information even in the face of more complex modulations. Besides, the detection performance of LS and MMSE algorithms is less affected by the modulation mode, but the detection accuracy is low. In addition, the complexity of the modulation mode has an obvious effect on the detection performance of DDLSD.

\begin{table}[h]
	\centering
	\small
	\caption{Comparison of BERs for different modulation modes.}\label{table:3}
	\begin{tabular}{c c c c}  
		\toprule[2pt]
		Algorithm & Modlation &SNR \tiny {($BER=10^{-1}$)}& SNR \tiny {($BER=10^{-2}$)}\\ \hline
		\multirow{2}{*}{LS} & QPSK & 19.5dB & 20dB+ \\
		& 16QAM & 20dB+ & 20dB+ \\ \hline
		\multirow{2}{*}{MMSE} & QPSK & 19.2dB & 20dB+ \\
		& 16QAM & 20dB+ & 20dB+ \\ \hline
		\multirow{2}{*}{DDLSD} & QPSK & 13dB & 19dB \\
		& 16QAM & 18dB & 20dB+ \\ 
		\bottomrule[2pt]
	\end{tabular}
\end{table}

\section{Conclusion}\label{SEC5}
In this paper, we proposed a signal detection scheme based on deep learning. Considering an OFDM system, we established a mathematical model of OFDM system. Based on data-driven, LSTM neural network was used for offline training of OFDM symbols, and the trained model was tested online. The proposed scheme is more robust to pilots, CP and modulation modes with the conventional detection methods. Simulation results demonstrate that the BER of the proposed method is obviously lower than traditional algorithms under different pilots, whether CP exists, and different modulation modes.

\section*{Acknowledgment}
This work was supported in part by the Natural Science Foundation of Jiangsu Province BK20200440, the Fundamental Research Funds for the Central Universities (No.1004-YAH20016, No.NT2020009).

\normalem
\bibliographystyle{IEEEtran}
\bibliography{ref}

\end{document}